\documentclass[fleqn,usenatbib]{mnras}

\usepackage{newtxtext,newtxmath}

\usepackage[T1]{fontenc}

\DeclareRobustCommand{\VAN}[3]{#2}
\let\VANthebibliography\thebibliography
\def\thebibliography{\DeclareRobustCommand{\VAN}[3]{##3}\VANthebibliography}

\usepackage{graphicx}	
\usepackage{amsmath}	
\usepackage[usenames, dvipsnames]{color}

\defcitealias{ferreira23}{F23}
\newcommand{\pfeat}{$p_{\rm{featured, GZ}}$}

\newcommand{\rjs}[1]{{\color{black}{{#1}}}}

\title[Featureless discs in JWST CEERS]{Galaxy Zoo JWST: Up to 75\% of discs are featureless at $3<z<7$}

\author[R. J. Smethurst et al.]{R. J. Smethurst,$^{1}$\thanks{E-mail: rebecca.smethurst@physics.ox.ac.uk} B. D. Simmons,$^{2}$ T. Géron,$^{3}$ H. Dickinson,$^{4}$ L. Fortson,$^{5,6}$ I. L. Garland,$^{7,2}$  S. Kruk,$^{8}$ \newauthor S. M. Jewell,$^{1}$ C. J. Lintott,$^{1}$ J. S. Makechemu,$^{2}$ K. B. Mantha,$^{5,6}$ K. L. Masters,$^{9}$ D. O'Ryan,$^{10}$ H. Roberts,$^{5,6}$ \newauthor M. R. Thorne,$^{2}$ M. Walmsley,$^{3}$ M. Calabr\`o,$^{11}$ B. Holwerda,$^{12}$ J. S. Kartaltepe,$^{13}$ A. M. Koekemoer,$^{14}$ \newauthor Y. Lyu,$^{15}$ R. Lucas,$^{16}$  F. Pacucci,$^{16,17}$ M. Tarrasse$^{15}$ \\
$^{1}$Department of Astrophysics, Denys Wilkinson Building, Keble Road, Oxford, OX1 3RH, UK\\
$^{2}$Department of Physics, Lancaster University, Lancaster LA1 4YB, UK\\
$^{3}$Dunlap Institute for Astronomy and Astrophysics, University of Toronto, 50 St. George Street, Toronto, ON M5S 3H4, Canada\\
$^{4}$School of Physical Sciences, The Open University, Milton Keynes, MK7 6AA, UK\\
$^{5}$School of Physics and Astronomy, University of Minnesota, Minneapolis, Minnesota, 55455, USA\\
$^{6}$Minnesota Institute for Astrophysics, University of Minnesota, Minneapolis, Minnesota, 55455, USA\\
$^{7}$Department of Theoretical Physics and Astrophysics, Faculty of Science, Masaryk University, Kotl\'{a}\v{r}sk\'{a} 2, Brno, 611 37, Czech Republic\\
$^{8}$European Space Agency (ESA), European Space Astronomy Centre (ESAC), Camino Bajo del Castillo s/n, 28692, Villaneuva de la Cañada, Madrid, Spain\\
$^{9}$Departments of Physics and Astronomy, Haverford College, Lancaster Avenue, Ardmore, PA 19041 USA\\
$^{10}$Department of Astrophysics, Centro de Astrobiología, INTA-CSIC, Camino Bajo del Castillo, s/n, 28692, Villanueva de la Cañada, Madrid, Spain\\
$^{11}$INAF Osservatorio Astronomico di Roma, Via Frascati 33, 00078 Monte Porzio Catone, Rome, Italy\\
$^{12}$University of Louisville, Department of Physics and Astronomy, 102 Natural Sciences Building, Louisville KY 40292 \\
$^{13}$Laboratory for Multiwavelength Astrophysics, School of Physics and Astronomy, Rochester Institute of Technology, Rochester, NY, 14623, USA\\
$^{14}$Space Telescope Science Institute, 3700 San Martin Drive, Baltimore, MD 21218, USA\\
$^{15}$Université Paris-Saclay, Université Paris Cité, CEA, CNRS, AIM, 91191, Gif-sur-Yvette, France\\
$^{16}$Center for Astrophysics $\vert$ Harvard \& Smithsonian, 60 Garden St, Cambridge, MA 02138, USA\\
$^{17}$Black Hole Initiative, Harvard University, 20 Garden St, Cambridge, MA 02138, USA\\
}

\date{Accepted 2025 March 26. Received 2025 March 26; in original form 2025 January 21}
\pubyear{2024}

\begin{document}
\label{firstpage}
\pagerange{\pageref{firstpage}--\pageref{lastpage}}
\maketitle

\begin{abstract}

We have not yet observed the epoch at which disc galaxies emerge in the Universe. While high-$z$ measurements of large-scale features such as bars and spiral arms trace the evolution of disc galaxies, such methods cannot directly quantify featureless discs in the early Universe. Here we identify a substantial population of apparently featureless disc galaxies in the Cosmic Evolution Early Release Science (CEERS) survey by combining quantitative visual morphologies of $\sim 7,000$ galaxies from the Galaxy Zoo JWST CEERS project with a public catalogue of expert visual and parametric morphologies.
While the highest-redshift featured disc we identify is at $z_{\rm{phot}}=5.5$, the highest-redshift featureless disc we identify is at $z_{\rm{phot}}=7.4$.  The distribution of S\'ersic indices for these featureless systems suggests that they truly are dynamically cold: disc-dominated systems have existed since at least $z\sim 7.4$. We place upper limits on the featureless disc fraction as a function of redshift, and show that up to $75\%$ of discs are featureless at $3.0<z<7.4$. This is a conservative limit assuming all galaxies in the sample truly lack features. With further consideration of redshift effects and observational constraints, we find the featureless disc fraction in CEERS imaging at these redshifts is more likely $\sim29-38\%$. We hypothesise that the apparent lack of features in a third of high-redshift discs is due to a higher gas fraction in the early Universe, which allows the discs to be resistant to buckling and instabilities.
\end{abstract}




\section{Introduction}
Classifying the morphologies of galaxies is crucial to our understanding of galaxy evolution as whole. The morphology encodes the history of each galaxy and is indicative of its dynamical state; with dispersion supported systems having undergone a redistribution of angular momentum (often via major mergers), whereas rotationally supported dynamically cold galaxies develop a range of easily identifiable features (such as bars, spiral arms, rings). Such features are a fossil record of the disc instabilities and interactions the galaxy has experienced. This dichotomy between dispersion and rotationally supported systems at low redshift is clear \citep{Nair2010a, Lintott2008, Willett2013}, with galaxies falling into the distinct morphological categories of the Hubble Sequence \citep{Hubble1926}. However, at higher redshifts in Hubble Space Telescope (HST) imaging, galaxies become more peculiar in morphology, with clumpier morphologies dominating \citep{Elmegreen2005, Simmons2017, Tohill2024}. 

Studying the change in morphologies over cosmic time therefore allows the epoch of the emergence of the Hubble Sequence to be determined. In particular, an ongoing open question is how and when hot gas discs in the early Universe dynamically settled into the strongly rotation-dominated structures that we see today? Answering such a question has many implications for our understanding of the role of galaxy and halo mergers, disc instabilities, and the gas and baryon fractions of galaxies in the early Universe. 

Previous observational studies with HST have attempted to determine the maximum redshift at which it is possible to detect the hallmark features of dynamically cold disc galaxies, such as bars \citep[e.g.][$z\sim2$]{Simmons2014} and spiral features \citep[e.g.][$z\sim2.5-3$]{MB2022}. The observational consensus from studies using HST imaging was that $z\sim2$ appeared to be the main period of disc assembly, during which rotationally supported galaxies begin to dominate the galaxy population \citep{Conselice2011, Kassin2012, Buitrago2014, Simons2017, Simmons2017, Costantin2022}. However, using HST, the field has been limited to studying the rest-frame optical morphology of galaxies at redshifts $z<2.8$ (using the longest wavelength Wide Field Camera 3 filter F160W). Using the James Webb Space Telescope (JWST), rest-frame optical images can now be obtained for galaxies out to $z\sim8$ using the F444W filter with the Near-Infrared Camera (NIRCam), and at higher redshifts with the Mid-Infrared Instrument (MIRI) at lower resolution. In addition, JWST's primary mirror is more than 2.5 times the diameter of HST's, meaning that deeper imaging is possible. These advancements have allowed the identification of morphological features in JWST imaging at higher redshifts than ever before, such as spiral arms \citep[out to $z\sim4$, e.g.][]{Wu2023, Kuhn2024}, bars \citep[out to $z\sim4$, e.g.][]{Guo2023, Constantin2023, LeConte2024} and clumps \citep[out to $z\sim8$, e.g][]{Kalita2024,Tohill2024}. 


However, there is still disagreement in the literature of cosmological volume simulations over the epoch of disc formation, or ``settling''. In current cosmological models employing $\Lambda$CDM, the gravitational collapse of small perturbations in the distribution of matter eventually leads to galaxy formation \citep{silk2012}. What follows is a mix of hierarchical structure formation through halo mergers and smooth accretion leading to the prediction that high redshift galaxies are compact, clumpy, and peculiar in morphology \citep{Cen2014, Shimizu2014}. However, since angular momentum will also be present as a result of tidal torques, then a rotating gas disc should form upon gravitational collapse \citep{fall80, barnes87, fall2002}. 

However, cosmological volume simulations only predict late-time evolution of disc galaxies in their simulated universes, with most studies struggling to produce disc galaxies earlier than $z\sim4$. For example, in the Millennium simulation, \cite{Parry2009} find a median formation epoch for spirals/S0s in the range $1.18 < z < 1.38$, with a maximum formation redshift of $z\sim3-4$. In the Illustris TNG50 simulations \cite{Pillepich2019} find an average V/$\sigma < 3$ for galaxies at $z>3$, while \cite{Semenov2024b} find a range of disc spin-up times (a proxy for disc formation time) of $2<z<5$ for Milky Way analogues. However, in the Horizon-AGN simulation, \cite{Dubois2016} track V/$\sigma$ evolution (a measure of the ratio between ordered and turbulent motion in a galaxy), showing rotation dominated galaxies with high V/$\sigma$ at $1~\rm{Gyr}$ ($z\sim6$) into the Universe's lifetime. \rjs{This lack of discs at high-z in simulations is perhaps unsurprising given the computational constraints on volume and resolution  which restrict the dynamic range in overdensities probed; \cite{lovell2021} have attempted to overcome this restriction with the FLARES simulation by employing a novel weighting scheme, but did not consider disc fractions in that work.}

In addition, zoom-in simulations \rjs{(which draw from the above parent cosmological volume simulations with limited dynamic ranges)} also struggle to form disc galaxies at early times. For example, \cite{Ceverino2017} show with the AGORA simulation that while a galaxy can evolve into a disc in just $0.5~\rm{Gyr}$, such galaxies are only in place by $z\sim1.2$. Using the VINTERGATAN simulation, \cite{Agertz2021} showed that for a Milky Way analogue galaxy, disc formation occurred between $1.3<z<3.5$. Similarly, in the FIRE2 simulations \cite{McCluskey2024} study the kinematic evolution of 14 Milky Way mass galaxies and find a range of disc settling epochs from $z \approx0.4 - 1$, with the earliest time of disc onset at $z\approx2$. 

However, note that \cite{Wilkinson2023, JBH2024} warn that simulated galaxies within halos with  $N<10^6$ dark matter particles are susceptible to spurious morphological evolution. \rjs{Similarly, \cite{ludlow2021} find that dark matter halos with $N<10^6$ particles will include spurious collisional heating of stars near the stellar half-mass radius such that their vertical velocity dispersion increases by 10\% of the halo’s virial velocity in roughly one Hubble time.} This was also discussed by \cite{Fujii2011} who state that $N\gtrsim10^6$ particles in discs are necessary in simulations to avoid artificial heating through close encounters with massive particles, which can lead to rounder spheroidal systems which loose their rotational support. In addition, \cite{Fujii2018} demonstrated how simulation size, $N$, is coupled to bar formation timescales in simulated disc galaxies. It is therefore unsurprising that there are some zoom-in simulations which are able to predict the existence of discs at early epochs. Using the SERRA simulation, \citet{Kohandel2024} found that dynamically cold discs are prevalent in the early Universe, with massive galaxies ($M_* > 10^{10}~M_{\odot}$) with $V/\sigma > 10$ at $z > 4$ maintaining cold discs for over 10 galaxy orbital periods. Critically, \citet{Kohandel2024} find no evolution in the average $V/\sigma$ with redshift out to $z=9$. This is in agreement with the work of \cite{Feng2015}, who find that kinematically classified discs (with $V/\sigma>4$) make up $70\%$ of the galaxy population at $z=8-10$ in the BlueTides hydrodynamical simulation, but that crucially these systems do not have features (see their Figure 1). 

\begin{figure*}
\includegraphics[width=\textwidth]{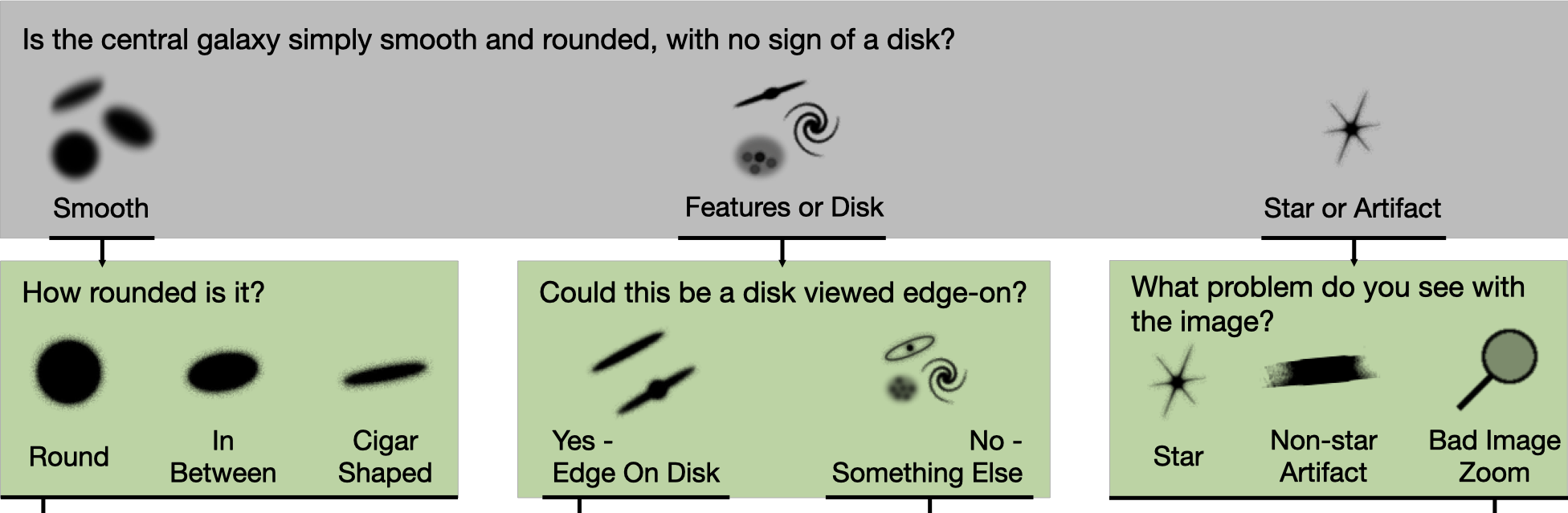}
\caption{The first two questions in the GZ CEERS decision tree, reproduced from the full version in \protect\cite{masters24prep}.}
\label{fig:tree}
\end{figure*}

If a galaxy is absent of features such as spiral arms and/or a bar, the typical assumption is that it is dynamically hot and therefore must have an elliptical or lenticular morphology; however, it is also possible for a galaxy to be dynamically cold with a smooth, featureless disc (such galaxies have been observed at low-redshift, e.g. \citealt{FM2018}). For galaxies to be truly featureless, the galaxy must be free of processes which induce perturbations in the stellar disc, whether density waves, disc instabilities, tidal interactions, or otherwise. \citealt{Elmegreen2013} showed how exponential discs develop from stellar scattering alone, and no other perturbations are required. Gas rich systems in particular are more resistant to buckling and instabilities \citep{Lokas2020}, with the high redshift Universe thought to have higher molecular gas fractions \citep[e.g.][]{Wang2022, Heintz2023}. In recent simulations using NEXUS, \cite{JBH2023,JBH2024} have shown how gas fraction is secondary to the baryonic vs. dark matter fraction within the disc, $f_{\rm{disc}}$, with relation to the formation epoch of bar structures. It is only when baryons dominate over dark matter, i.e. $f_{\rm{disc}}>0.5$, that bars form in discs after $0.7~\rm{Gyr}$ \citep{JBH2024}; prior to that the discs are featureless. Therefore, previous studies which identify solely large-scale features in an attempt to characterise the epoch of the emergence of the Hubble Sequence which do not account for the possibility of a featureless disc, will underestimate the redshift at which discs emerge. 

However, without the features that identify a galaxy as rotationally supported (e.g. bars, spiral arms), identifying a featureless galaxy as a dynamically cold system with imaging data alone is not trivial. The light profile of a galaxy is thought to trace whether a galaxy is rotational or dispersion dominated, resulting in the possibility of both a visual and parametric classification of a disc, reducing the need for spectral follow-up to trace the stellar kinematics. Classifying galaxy morphologies however, is not trivial. Due to the large size of current datasets, the field has turned to automated morphological classifications which fall into two categories. First, fitted morphologies (e.g. parametric morphologies such as Sérsic index, or non-parametric morphologies such as concentration and asymmetry), which have been a mainstay of the field for decades \citep{Abraham1994, Shade1995, Abraham1996, Conselice2003, Lotz2004, Tohill2021, Lazar2024}. However with increasing redshift, fitted morphologies become less-informative due to resolution limitations. In addition, fitted morphologies reduce the complex information in images to single numbers; while useful, these parameters lack the complexity that a visual classification can capture. Secondly, machine learning has advanced this field significantly \citep{Dieleman2015, HuertasCompany2015, Hocking2015, Khan2019, DS2019, Cheng2020, Tang2022, Ferreira2020, Ghosh2023, Bom2024}, 
however, large training sets (with $\sim10^5$ expert classifications; \citealt{Dieleman2015}) do not yet exist for JWST imaging. 
Therefore, a visual morphological classification of a galaxy from an image still provides the most informative and precise morphologies possible (at least, given the data currently available for JWST). For smaller datasets this is achievable for a science collaboration, however with increasing dataset size this becomes unfeasible.

The Galaxy Zoo collaboration has spent nearly two decades running online citizen science projects which ask the public to classify images from large galaxy surveys such as the Sloan Digital Sky Survey \citep[SDSS]{Lintott2009, Willett2013}, the  Dark Energy Camera Legacy Survey \citep[DECaLS]{Walmsley2022}, the 
Dark Energy Spectroscopic Instrument \citep[DESI]{Walmsley2023}, and HST Cosmic Assembly Near-infrared Deep Extragalactic Legacy Survey \citep[CANDELS]{Simmons2017}. Recently, \cite{Walmsley2020} have implemented a dual approach with \mbox{DECaLS} imaging, using active learning to focus the volunteer effort on the galaxies which, if labeled, would be most informative for training their CNN. Similarly, to prepare for the size of planned galaxy surveys with JWST, the Galaxy Zoo collaboration has completed a pilot study using public JWST CEERS \citep{Finkelstein2023} NIRCam imaging of $\sim7000$ galaxies \citep{masters24prep}. In the rest of this paper we will refer to this project as GZ CEERS.

GZ CEERS used a classification tree system to guide volunteers through a morphological classification, the top of which is shown in Figure~\ref{fig:tree} (for the full decision tree please see \citealt{masters24prep}). The first question presented to volunteers was ``Is the central galaxy simply smooth and rounded, with no sign of a disc?''. The response options were ``smooth'', ``features or disc'', and ``star or artifact''. This resulted in a vote fraction of \pfeat~for each galaxy classified, i.e. the number of people who picked ``features or disc'' divided by the total number of people who responded to the question. Previous Galaxy Zoo projects have demonstrated how volunteers are mostly influenced by the presence of features in a galaxy, rather than the presence of a disc, when determining their response to this first question \rjs{\citep[e.g. see][]{Simmons2017, Willett2017}}. Featureless disc galaxies, lenticulars, and S0 galaxies have always been classified by volunteers in previous GZ projects as ``smooth'' \citep{Willett2013, Simmons2017, Walmsley2020, Walmsley2023}. For low-redshift Galaxy Zoo projects, a bimodal distribution in \pfeat~is typically seen \citep{Willett2013, Walmsley2022}, with \pfeat~ allowing for a clean selection of a sample of featured disc galaxies with high completeness. However, \cite{masters24prep} reported a one-sided distribution for GZ CEERS with a peak at \pfeat$\sim0.25$. Therefore, this initial question of the decision tree in the pilot GZ CEERS project does not allow for the selection of a sample of disc galaxies with high completeness like at low-redshift. 

However, given the high disc fractions of $\sim40-60\%$ reported by other studies that have visual or automated classifications of the same CEERS field \citep[e.g.][]{ferreira23,HC2024,Kartaltepe2023,Tohill2024}, the low average \pfeat~in GZ CEERS suggests there is a large number of featureless discs in the CEERS field. The GZ CEERS classifications in combination with expert visual or parametric morphological classifications, are therefore incredibly useful for identifying a sample of this forgotten population of featureless disc galaxies at high redshift. This prevalence of featureless discs is unsurprising given the high redshift range probed by CEERS, where observational biases prevail, however, given the predictions from simulations discussed above, in this study we investigate the possibility of whether these galaxies are truly featureless.

In this paper we investigate the fraction of discs that appear featureless as a function of redshift, discuss whether they are truly dynamically cold discs, and if so, why they lack the features typical of disc galaxies. We describe the data sources and sample selection in Section \ref{sec:data}. Our results are shown in Section \ref{sec:results}, we discuss the implications of these results in Section \ref{sec:discuss}, and we conclude in Section \ref{sec:conclusions}. In the rest of this work we adopt the Planck 2015 \citep{planck16} cosmological parameters with $(\Omega_m, \Omega_{\Lambda}, h) = (0.31, 0.69, 0.68)$.

\section{Data \& Methods}
\label{sec:data}

\subsection{Galaxy Zoo CEERS}

The GZ CEERS project saw volunteers classifying images derived from the CEERS data release 0.5 \citep{Bagley23} of NIRCam \citep{Rieke23} imaging in the F115W, F150W, F200W, F277W, F356W and F444W broadband filters, plus the F410M narrow band filter. Volunteers were shown chromatically mapped colour images (publicly released by the CEERS team), allowing the use of information across many bands in allocating morphological labels. For more details on the GZ CEERS project see \cite{masters24prep}. During the period of June to November 2023, $7679$ JWST CEERS cutout images were shown to volunteers on the Galaxy Zoo website, which resulted in $311,413$ unique classifications collected from $3496$ unique Zooniverse volunteers logged in with their username ($6957$ classifiers including those not logged in). \rjs{The median number of classifications per image was $40$, with a minimum of $39$ and maximum of $89$. The median number of classifications per volunteer was $12$, with a minimum of $1$ and a maximum of $6945$ classifications}. Note that the initial catalogue creation was intended only for object detection prior to the availability of a science-ready catalogue, and only minimal effort was put initially into duplicate removal, so some galaxies were certainly shown in multiple images. Duplicates were then removed in post-production, resulting in $6689$ images with classifications. 

\begin{figure}
    \includegraphics[width=0.5\textwidth]{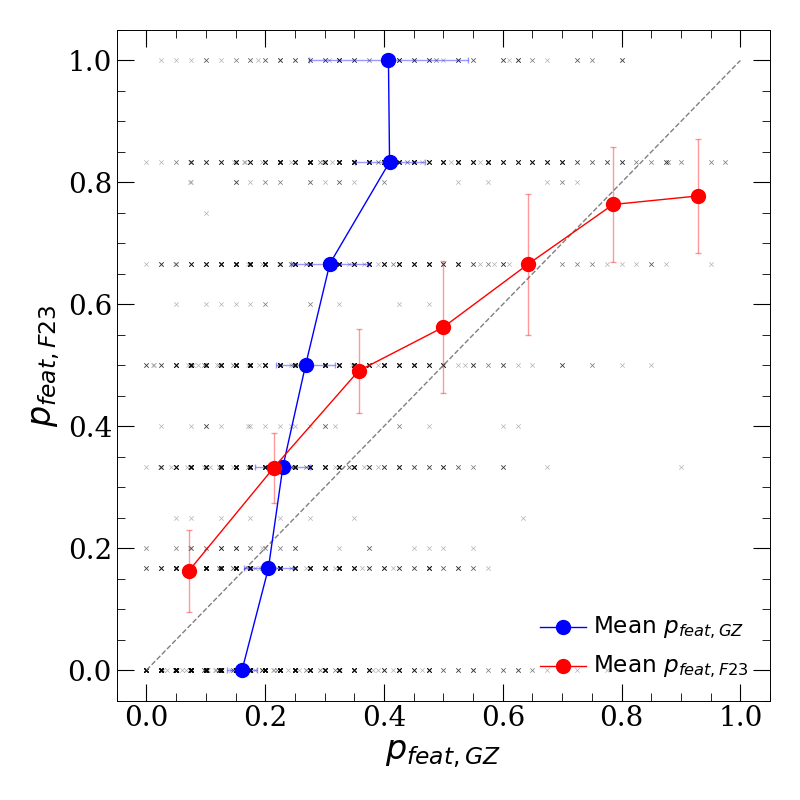}
    \caption{The GZ CEERS featured vote fraction, $p_{\rm{feat, GZ}}$, against the expert featured vote fractions from \protect\citetalias{ferreira23}, $p_{\rm{feat, F23}}$, shown with the grey crosses (an increase in black saturation indicates an overlap of many objects). \rjs{Note that if one/two authors classified a galaxy as unclassifiable, $p_{\rm{feat, F23}}$ were calculated with only five/four author votes by \protect\citetalias{ferreira23}}.  The dashed grey line shows a 1:1 comparison. The red circles show the mean disc vote fraction from \protect\citetalias{ferreira23} in equal sized bins of GZ CEERS featured vote fraction (i.e. binned on the x-axis). The blue circles show the mean GZ CEERS featured vote fraction in equal sized bins of disc vote fraction from \protect\citetalias{ferreira23} (i.e. binned on the y-axis). The uncertainties on the mean values are the range of means found during 1000 bootstrap iterations recalculating with a random selection of $90\%$ of the sample in each bin. This figure shows how the GZ volunteers are less likely to classify a galaxy as featured than \citetalias{ferreira23}, and are more ``pessimistic" with their classifications.}
    \label{fig:smoothcompare}
\end{figure}

We first removed those images which are classified as ``star or artifact'' by the volunteers from the overall sample and use a threshold of $p_{\rm{artifact}}<0.5$ to select classifiable galaxies. This threshold results in the number of galaxies that were classifiable by GZ CEERS $\rm{N}_{\rm{GZ}}=5734$. The decision of what threshold on the vote fraction to use to obtain a clean sample of a subset of galaxies with a given morphology is not trivial \citep[see more detailed discussion from previous Galaxy Zoo projects e.g.][]{Lintott2009, Willett2013, Walmsley2022, Walmsley2023}. Three authors (TG, IG, DO) independently visually inspected the NIRCam imaging alongside the user vote fractions to determine the threshold at which a clean sample was obtained for a given question in the decision tree (see Figure~\ref{fig:tree}). Emphasis was placed on purity over completeness in the decision of thresholds. To select featured galaxies we therefore adopt a vote fraction \pfeat$>0.3$. This results in $\rm{N}_{\rm{feat,GZ}}=1912$ featured galaxies.


\subsection{Redshifts}
Redshifts were obtained from the HST CANDELS project \citep{Grogin2011, Koekemoer2011}, by cross-matching to the CEERS data on sky co-ordinates. \rjs{We note that at the time of preparation and submission of this manuscript, finalised CEERS photometric redshifts were not yet available. The publically available published CANDELS redshifts} are a mix of photometric redshifts measured from multi-wavelength broadband HST imaging, and spectroscopic redshifts where available \citep[see][for details on redshifts]{Kodra2023}. All of the GZ CEERS galaxies were matched to HST CANDELS within 5 arcsecs radius.  $87\%$ of the GZ CEERS sample have photometric redshifts from CANDELS. 

\subsection{Comparison sample from \protect\cite{ferreira23}}\label{subsec:f23}

\begin{figure*}
    \includegraphics[width=\textwidth]{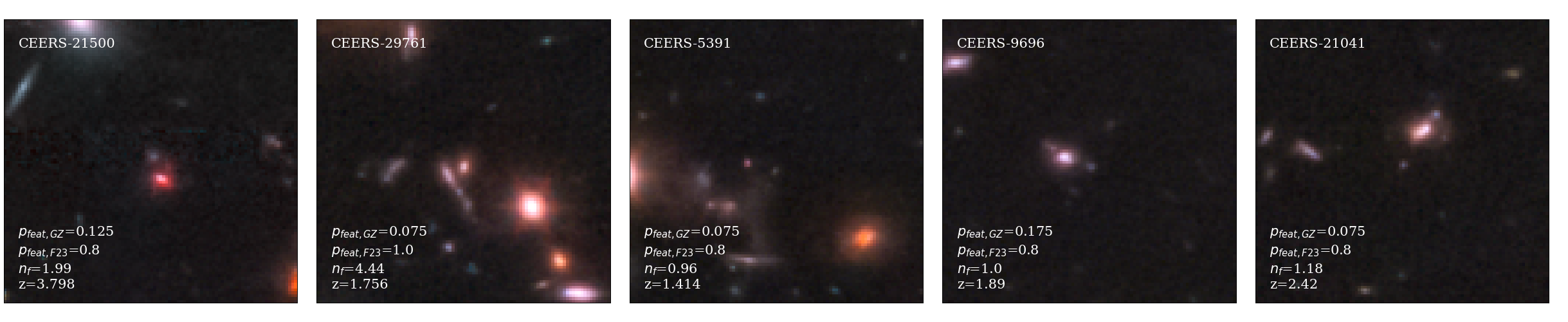}
    \caption{Example $rgb$ JWST CEERS images visually identified by \protect\citetalias{ferreira23} as having a high featured vote fraction ($p_{\rm{feat,F23}}>0.8$) but with low featured vote fractions in GZ CEERS ($p_{\rm{feat,GZ}}<0.2$). In the top left corner of each image we provide the CEERS ID of the galaxy. In the bottom left corner of each image we provide the GZ featured vote fraction, $p_{\rm{feat,GZ}}$, the \protect\citetalias{ferreira23} featured vote fraction, $p_{\rm{feat,F23}}$, the Sérsic index fit by \protect\citetalias{ferreira23}, $\rm{n}_F$, and the HST CANDELS redshift, $z$, for each galaxy. In total, 62 galaxies fall into this category. Note that these are the exact images which were shown to GZ CEERS volunteers during classification; volunteers were asked to classify the galaxy in the centre of the image.}
    \label{fig:GZnotF23feat}
\end{figure*}

In order to select a sample of featureless discs we must combine the Galaxy Zoo featured classifications with expert morphological disc classifications. Here we use the visual and parametric morphologies of $4265$ galaxies with CEERS imaging from \citet[][hereafter F23]{ferreira23}. In that study, 6 experts visually classified all sources into 5 categories: unclassifiable, point sources, peculiar, spheroids, and discs. \citetalias{ferreira23} identified discs as those galaxies where more than half of the authors classified the galaxy as a disc \rjs{(note that if one author classified a galaxy as unclassifiable, the resultant classification fractions for spheroids and discs were calculated with only five author votes)}. In addition \citetalias{ferreira23} obtained both parametric and non-parametric morphologies using the \textsc{MORFOMETRYKA} algorithm \citep{ferrari15}; in particular for this work they fitted Sérsic profiles to each galaxy in all the filter images (but only report results for the band closest to the rest-frame optical at $\lambda=5000-7000$\AA). \citetalias{ferreira23} first performed one-dimensional Sérsic fits on the luminosity profile of each galaxy, which they then used as input for a two-dimensional Sérsic fit done with the galaxy and the PSF images. Note that the authors of \citetalias{ferreira23} were not shown the Sérsic index fit during their visual inspection. 

We cross-matched the GZ CEERS sample with the \citetalias{ferreira23} sample, resulting in $3069$ galaxies within a 5 arcsec separation which were morphologically classified by both studies. Note that \citetalias{ferreira23} also visually classify all of their galaxies as either smooth or featured, but do not investigate this further in their original study. Here, we compare their featured vote fraction with the GZ featured vote fraction in Figure~\ref{fig:smoothcompare}. The red circles in Figure~\ref{fig:smoothcompare} show the mean featured vote fraction from \citetalias{ferreira23} in equal sized bins of GZ CEERS featured vote fraction; the mean roughly traces the 1:1 line between the two vote fractions, although \citetalias{ferreira23} tend to be more likely to classify a galaxy as featured, whereas GZ volunteers are more ``pessimistic''. Similarly, the blue circles show the mean GZ CEERS featured vote fraction in equal sized bins of featured vote fraction from \citetalias{ferreira23}, which do not trace the 1:1 line, once again demonstrating that GZ volunteers are less likely to classify a galaxy as featured than \citetalias{ferreira23}. While there are not any galaxies which the GZ volunteers classify as featured but \citetalias{ferreira23} do not (bottom right corner of Figure~\ref{fig:smoothcompare}), there are $62$ galaxies which GZ volunteers do not classify as featured ($p_{\rm{feat,GZ}}<0.2$; top left corner of Figure~\ref{fig:smoothcompare}), but \citetalias{ferreira23} do ($p_{\rm{feat,F23}}>0.8$). Examples of such galaxies are shown in Figure~\ref{fig:GZnotF23feat}. 

\begin{figure*}
\includegraphics[width=\textwidth]{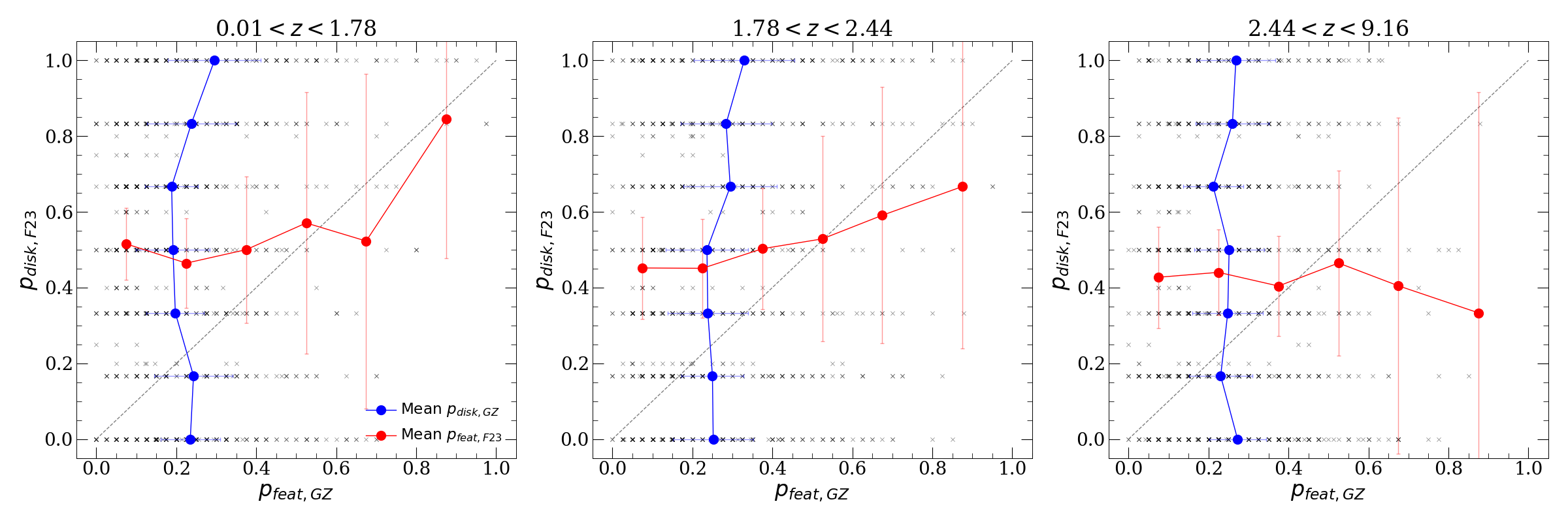}
    \caption{The GZ CEERS featured vote fraction, $p_{\rm{feat, GZ}}$, against the disc vote fractions from \protect\citetalias{ferreira23}, $p_{\rm{disc, F23}}$, shown in each panel with the black crosses. We split the sample into low, medium and high redshift bins with a third of the sample in each bin. \rjs{Note that if one/two authors classified a galaxy as unclassifiable, $p_{\rm{disk, F23}}$ were calculated with only five/four author votes by \protect\citetalias{ferreira23}}. In each panel the dashed grey line shows a 1:1 comparison. The red circles show the mean disc vote fraction from \protect\citetalias{ferreira23} in equal sized bins of GZ CEERS featured vote fraction (i.e. binned on the x-axis). The blue circles show the mean GZ CEERS featured vote fraction in equal sized bins of disc vote fraction from \protect\citetalias{ferreira23} (i.e. binned on the y-axis). The uncertainties on the mean values are the range of means found during 1000 bootstrap iterations recalculating with a random selection of $90\%$ of the sample in each bin. Note that the uncertainties on bins with high $p_{\rm{feat, GZ}}$ are larger due to small number statistics; as discussed in Section~\ref{subsec:f23}, GZ volunteers are less likely than \citetalias{ferreira23} to classify a galaxy as featured.}
    \label{fig:comparefeatdisc}
\end{figure*}

\begin{figure*}
\includegraphics[width=\textwidth]{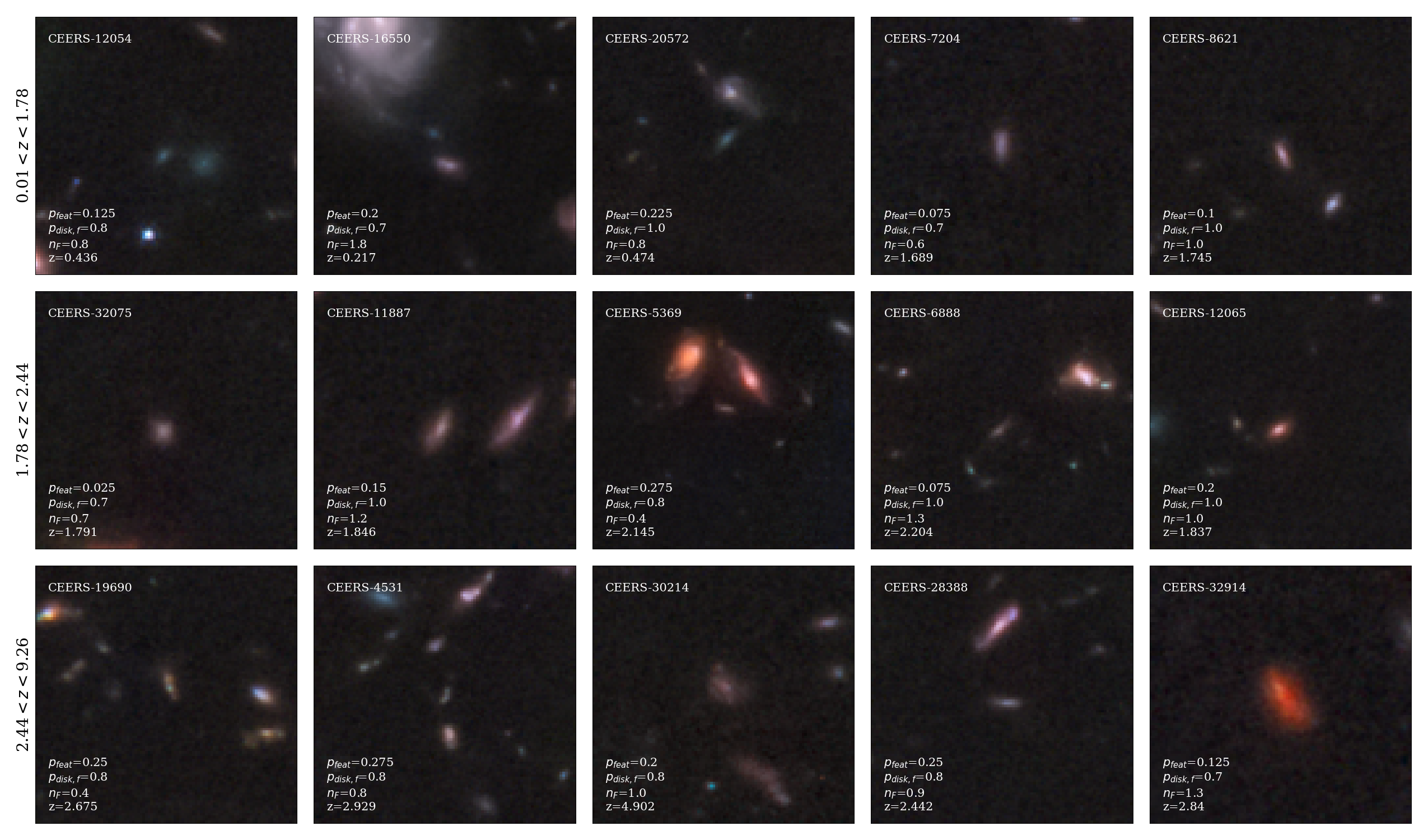}
    \caption{Example $rgb$ JWST CEERS images visually identified by \protect\citetalias{ferreira23} as a disc ($p_{\rm{disc,F23}}>0.5$) but with low featured vote fractions in GZ CEERS ($p_{\rm{feat,GZ}}<0.3$); aka apparently featureless discs. We show a representative sample at low redshift (top row; $0.01 < z < 1.78$), medium redshift (middle row; $1.78 < z< 2.44$), and high redshift (bottom row; $2.44 < z < 9.26$). In the top left corner of each image we provide the CEERS ID of the galaxy. In the bottom left corner of each image we provide the GZ featured vote fraction, $p_{\rm{feat, GZ}}$, the \protect\citetalias{ferreira23} disc vote fraction, $p_{\rm{disc, F23}}$, the Sérsic index fit by \protect\citetalias{ferreira23}, $\rm{n}_F$, and the HST CANDELS redshift, $z$, for each galaxy. The examples shown are representative of a sample of apparently featureless discs one can select by combining GZ CEERS with expert classifications and/or parametric morphologies from \protect\citetalias{ferreira23}. Note that these are the exact images which were shown to GZ CEERS volunteers during classification; volunteers were asked to classify the galaxy in the centre of the image.}
    \label{fig:lowpfeathighpdisc}
\end{figure*}

\begin{figure*}
\includegraphics[width=\textwidth]{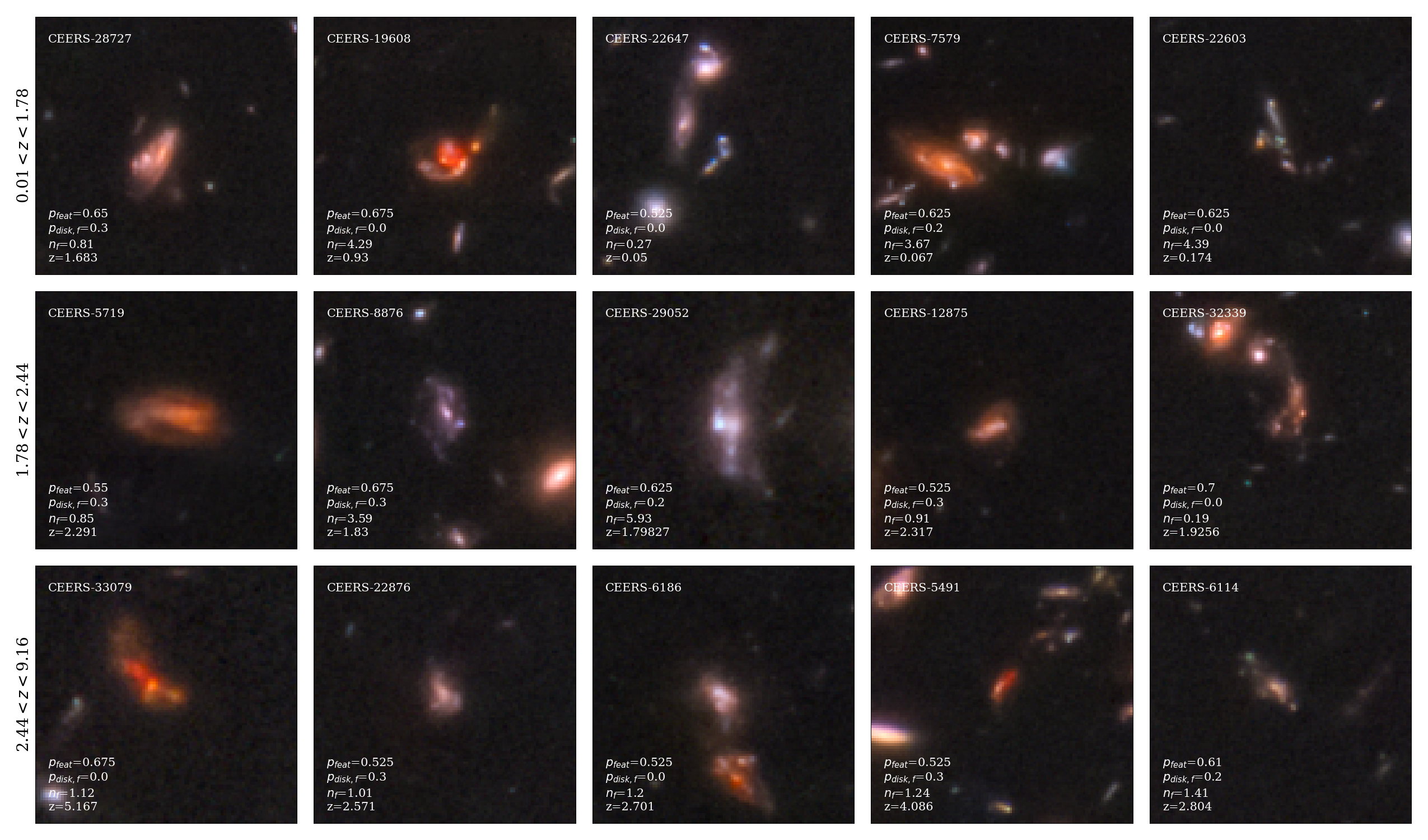}
    \caption{Example $rgb$ JWST CEERS images visually identified by \protect\citetalias{ferreira23} as not a disc ($p_{\rm{disc,F23}}<0.5$) but with high featured vote fractions in GZ CEERS ($p_{\rm{feat,GZ}}>0.5$). We show a representative sample at low redshift (top row; $0.01 < z < 1.78$), medium redshift (middle row; $1.78 < z< 2.44$), and high redshift (bottom row; $2.44 < z < 9.26$). In the top left corner of each image we provide the CEERS ID of the galaxy. In the bottom left corner of each image we provide the GZ featured vote fraction, $p_{\rm{feat, GZ}}$, the \protect\citetalias{ferreira23} disc vote fraction, $p_{\rm{disc,F23}}$, the Sérsic index fit by \protect\citetalias{ferreira23}, $\rm{n}_F$, and the HST CANDELS redshift, $z$, for each galaxy are shown. The examples shown are representative of a sample of featured irregular/clumpy galaxies one can select by combining GZ CEERS with expert classifications and/or parametric morphologies from \protect\citetalias{ferreira23}. Note that these are the exact images which were shown to GZ CEERS volunteers during classification; volunteers were asked to classify the galaxy in the centre of the image.}
    \label{fig:pfeatforvisualdiscs}
\end{figure*}

\begin{figure*}
\includegraphics[width=\textwidth]{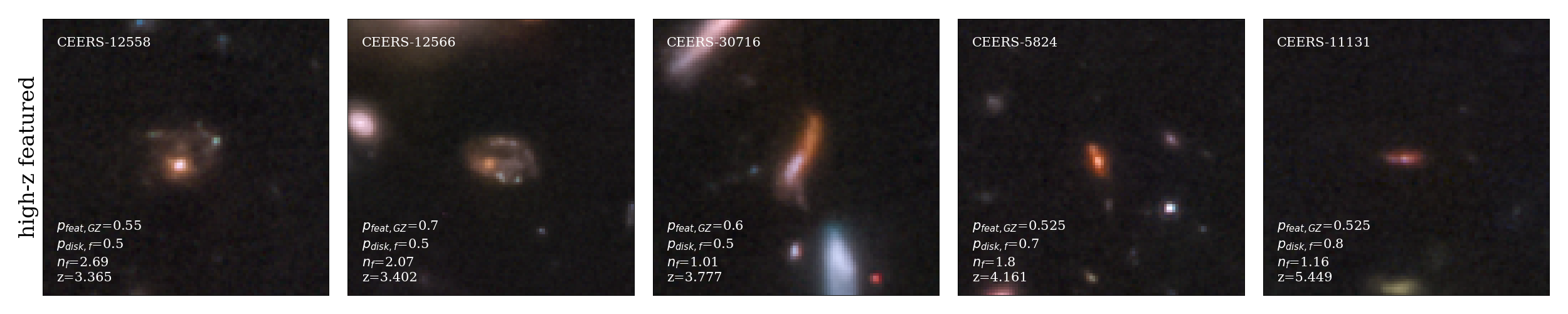}
\includegraphics[width=\textwidth]{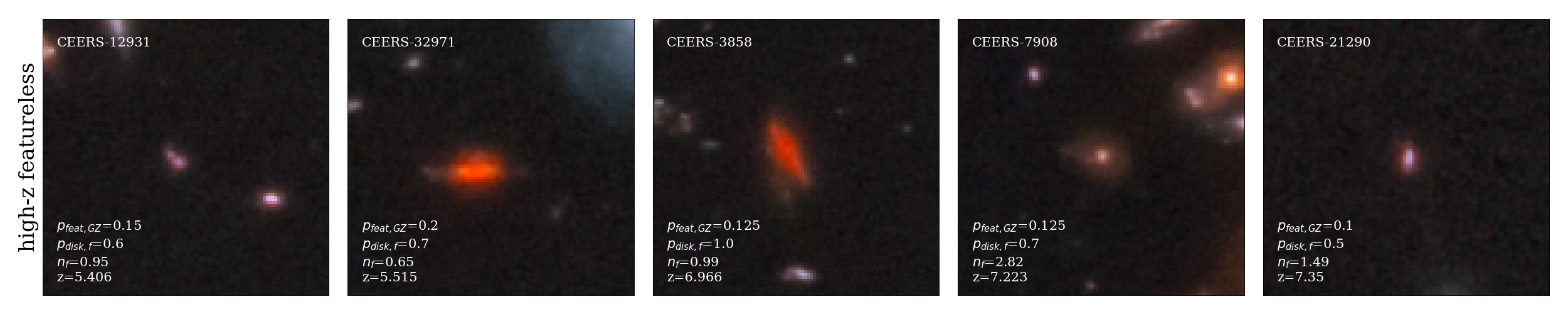}
    \caption{The highest redshift candidate featured (top row) and featureless (bottom row) discs found in GZ CEERS. In the top left corner of each image we provide the CEERS ID of the galaxy. In the bottom left corner of each image the GZ featured vote fraction, $p_{\rm{feat, GZ}}$, the \protect\citetalias{ferreira23} disc vote fraction, $p_{\rm{disc,F23}}$, the Sérsic index fit by \protect\citetalias{ferreira23}, $n$, and the HST CANDELS redshift, $z$, for each galaxy are shown. Note that these are the exact images which were shown to GZ CEERS volunteers during classification; volunteers were asked to classify the galaxy in the centre of the image.}
    \label{fig:highestz}
\end{figure*}

\begin{figure}
\includegraphics[width=0.49\textwidth]{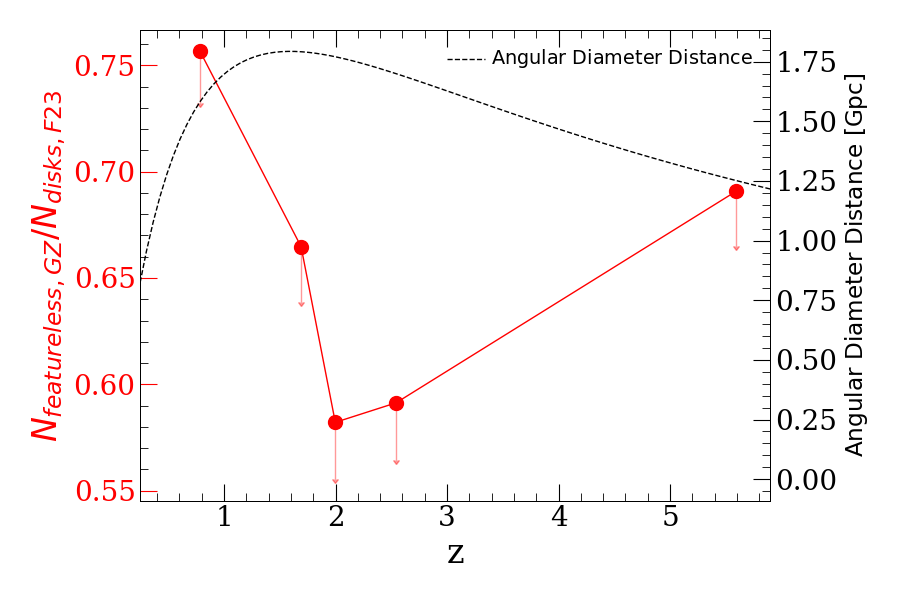}
    \caption{The upper limits on the fraction of apparently featureless discs identified by GZ CEERS as a function of redshift. Featureless discs are identified as those with GZ CEERS vote fractions, $p_{\rm{feat,GZ}}<0.3$. Whereas \protect\citetalias{ferreira23} visually identify discs as those galaxies with an expert disc vote fraction $p_{\rm{disc,F23}}>0.5$. \rjs{The angular diameter distance is also shown by the dashed line using the secondary y-axis on the right}. The data is split into redshift bins with the same number of galaxies per bin, and the points are plotted at the centre of the bins. Upper limit arrows are scaled to show the uncertainties on each calculated fraction.}
\label{fig:featurelesswithz}
\end{figure}

\begin{figure}
\includegraphics[width=0.49\textwidth]{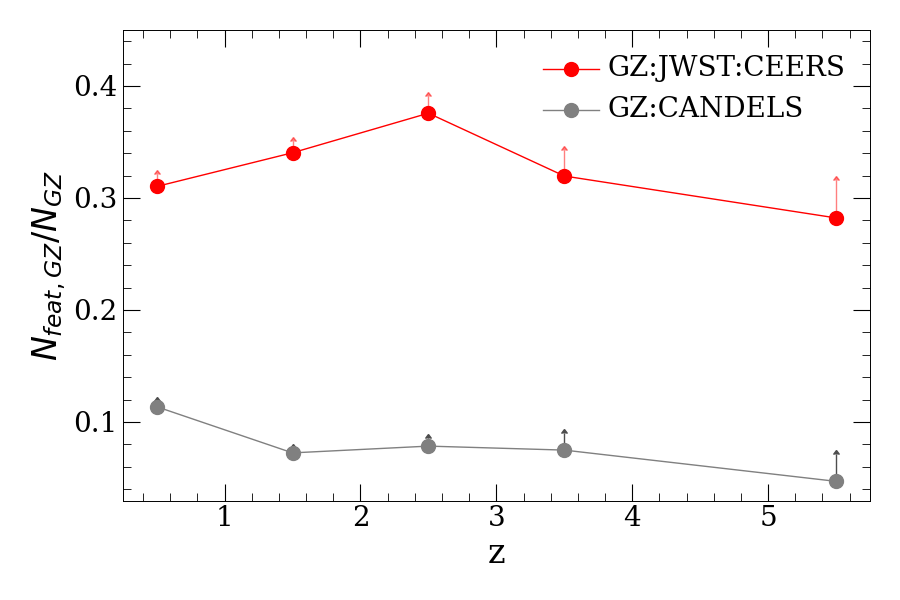}
\caption{The lower limit on the fraction of galaxies classified as featured in GZ CEERS, $\rm{N}_{\rm{feat, GZ}}/N_{\rm{GZ}}$, as a function of redshift, $z$, (red points). In grey, we also plot the equivalent fraction for galaxies in the GZ CANDELS project with images taken by HST (note there is no cross-over between the galaxies clasified for GZ CANDELS and GZ CEERS). Lower limit arrows are scaled to show the uncertainties on each calculated fraction. The fraction of featured galaxies in GZ CEERS increases with redshift from $\sim31\%$ at low-z to $\sim38\%$ at $2<z<3$, before decreasing to $\sim28\%$ at $z>5$. Whether this is a true redshift evolution, a result of the resolution limit of JWST, or due to the angular diameter distance turnover is discussed in Section~\ref{subsec:zlimit}.}
\label{fig:fractionwithz}
\end{figure}

\section{Results}\label{sec:results}

To better understand how the GZ CEERS featured vote fraction, \pfeat, relates to the expert disc vote fraction from \citetalias{ferreira23}, $p_{\rm{disc, F23}}$, we plot the two classifications against each other in Figure~\ref{fig:comparefeatdisc} and show the moving averages for 3 redshift bins. In the lowest redshift bin (left panel of Figure~\ref{fig:comparefeatdisc}), if the GZ CEERS volunteers identify a galaxy as featured, \pfeat$>0.3$, then \citetalias{ferreira23} are, on average, likely to have classified that galaxy as a disc, i.e. $p_{\rm{disc, F23}}>0.5$ (see red circles showing the mean disc vote fraction from \protect\citetalias{ferreira23} in equal sized bins of GZ CEERS featured vote fraction). However, with increasing redshift, this agreement dissipates (see right panel of Figure~\ref{fig:comparefeatdisc}). Conversely, if \citetalias{ferreira23} identified a galaxy as a disc, or otherwise, the average \pfeat~from GZ CEERS does not change, with volunteers equally likely to classify the galaxy as featureless or featured (see blue circles showing the mean GZ CEERS featured vote fraction in equal sized bins of disc vote fraction from \protect\citetalias{ferreira23}). This is consistent across all three redshift bins in each panel of Figure~\ref{fig:comparefeatdisc}.

To better understand the agreement and disagreement between vote fractions shown in Figure~\ref{fig:comparefeatdisc}, we performed a visual inspection of galaxies found in the upper left quadrants of the panels in Figure~\ref{fig:comparefeatdisc}, i.e. those visually identified by \protect\citetalias{ferreira23} as a disc ($p_{\rm{disc,F23}}>0.5$) but with low featured vote fractions in GZ CEERS (\pfeat$<0.3$). We identify $839$ such ``featureless" discs. A random sample of 15 of these galaxies is shown in Figure~\ref{fig:lowpfeathighpdisc}. In each panel we provide \pfeat, $p_{\rm{disc,F23}}$, the Sérsic fit from \citetalias{ferreira23}, $n_F$, and the HST CANDELS redshift, $z$, of each galaxy. A visual inspection confirms that these galaxies indeed appear to be featureless discs, with a median $n_F=1.16$ and a range of $0.2< n_F < 4.5$. 

Similarly, we then performed a visual inspection of galaxies found in the lower right quadrants of the panels in Figure~\ref{fig:comparefeatdisc}, i.e. visually identified by \protect\citetalias{ferreira23} as not a disc ($p_{\rm{disc,F23}}<0.5$) but with high featured vote fractions in GZ CEERS (\pfeat$>0.3$). We identify $433$ such galaxies. A random sample of 15 of these galaxies is shown in Figure~\ref{fig:pfeatforvisualdiscs}. In each panel we provide \pfeat, $p_{\rm{disc,F23}}$, the Sérsic fit from \citetalias{ferreira23}, $n_F$, and the HST CANDELS redshift, $z$, of each galaxy. A visual inspection of these galaxies shows a mix of peculiar and clumpy morphologies, with a median $n_F=1.6$ and a range of $0.2< n_F < 21.0$. 

In Figure~\ref{fig:highestz}, we then investigate the highest redshift featured discs (top row) and featureless discs (bottom row), of the two populations discussed above. The highest redshift apparently featureless disc we identify is CEERS-21290 at $z_{\rm{phot}}=7.35$ ($t_{\rm{lookback}}=13.1~\rm{Gyr}$) with Sérsic index $n=1.5$, and similarly, the highest redshift featured disc we identified is CEERS-11131 with both a disc and bulge feature at $z_{\rm{phot}}=5.5$ ($t_{\rm{lookback}}=12.8~\rm{Gyr}$) with Sérsic index $n=1.2$. While spectral follow-up will be required to confirm the redshifts of these systems, it is clear that any study of the epoch of the emergence of discs in the Universe's history must therefore include featureless discs, which can be identified out to much higher redshift than featured discs (e.g. the current highest redshift galaxies identified with bars or spiral arms are at $z\sim3-4$; \citealt{Wu2023, Guo2023, Kuhn2024}). 

We investigate this further by considering the change in the fraction of apparently featureless galaxies with redshift in Figure~\ref{fig:featurelesswithz}. The number of featureless galaxies decreases from $\sim76\%$ at $z\sim1$ to $\sim58\%$ at $z\sim2$, before increasing again to $\sim70\%$ at $z>3$. There are many observational biases at play here so the points plotted are upper limits; resolution effects will likely dominate, however we must also consider the turnover in the angular diameter distance, which should allow features to be more easily spotted at $z\sim2$ due to magnification effects caused by the expansion of the Universe; the drop in the apparently featureless fraction could therefore be due to an increase in the denominator, i.e. the number of discs identified by \citetalias{ferreira23}.

To test this hypothesis, we also look how the fraction of galaxies identified as featured in GZ CEERS, $\rm{N}_{\rm{feat,GZ}}/\rm{N}_{\rm{GZ}}$, changes with redshift, as shown in the top panel of Figure~\ref{fig:fractionwithz}. The fraction of featured galaxies increases with redshift from $\sim31\%$ at low-z, peaking at $\sim38\%$ at $2<z<3$, before decreasing to $\sim28\%$ at $z>5$. This suggests that the turnover in the angular diameter distance allows for easier feature classification at $z\sim2-3$, possibly explaining the drop in the apparent fraction of featureless discs seen in Figure~\ref{fig:featurelesswithz}. Otherwise, the fraction of featured galaxies identified by GZ volunteers is fairly flat with redshift,  within the uncertainties. This suggests that the change in the fraction of apparently featureless discs seen in Figure~\ref{fig:featurelesswithz} is not entirely driven by observational biases. 

In Figure~\ref{fig:fractionwithz}, we also show the fractions found in the GZ CANDELS project \citep{Simmons2017} which had volunteers classify images from HST surveys (including GOODS-S, UDS, and COSMOS). Note there is no overlap between the samples from GZ CEERS and GZ CANDELS, so we cannot directly compare featured vote fractions for specific galaxies. However, we can compare how the average vote fractions over the full samples changed with redshift in Figure~\ref{fig:fractionwithz} (note that \citealt{Simmons2017} also used a mix of photometric and spectroscopic redshifts, where available). There is a $\sim3-4$ fold increase in the number of featured galaxies identified with CEERS than with CANDELS, directly demonstrating the effect of JWST's improved resolution, sensitivity, and light collecting power over HST. 
Interestingly, the featured fraction shown in Figure~\ref{fig:fractionwithz} is lower than the disc fraction reported by other studies in JWST imaging at lower redshifts, however agrees with those found at higher redshifts of $z>5$. For example, \citetalias{ferreira23} find a disc fraction of $\sim45\%$ between $1.5<z<2$ and $\sim30\%$ between $4<z<8$, whereas \cite{Kartaltepe2023} find that discs make up $60\%$ of galaxies at $z=3$, dropping to $\sim30\%$ at $z=6-9$. The agreement at $z>5$ between the fractions reported by previous studies and found here with GZ CEERS, suggests that both the expert visual classifiers and the GZ volunteers are only able to identify discs if they are featured at higher redshift. However, the disagreement in the fractions at lower redshifts $z<5$, suggests that expert visual classifiers are \rjs{apparently} also able to identify featureless discs, which are not identified by volunteers on Galaxy Zoo. Comparing these findings in conjunction with the results presented above suggests a large fraction of disc galaxies in JWST imaging may be truly featureless. We discuss whether this is indeed the case below.

\section{Discussion}\label{sec:discuss}

\subsection{Are these truly featureless galaxies?}\label{subsec:zlimit}

\begin{figure*}
\includegraphics[width=\textwidth]{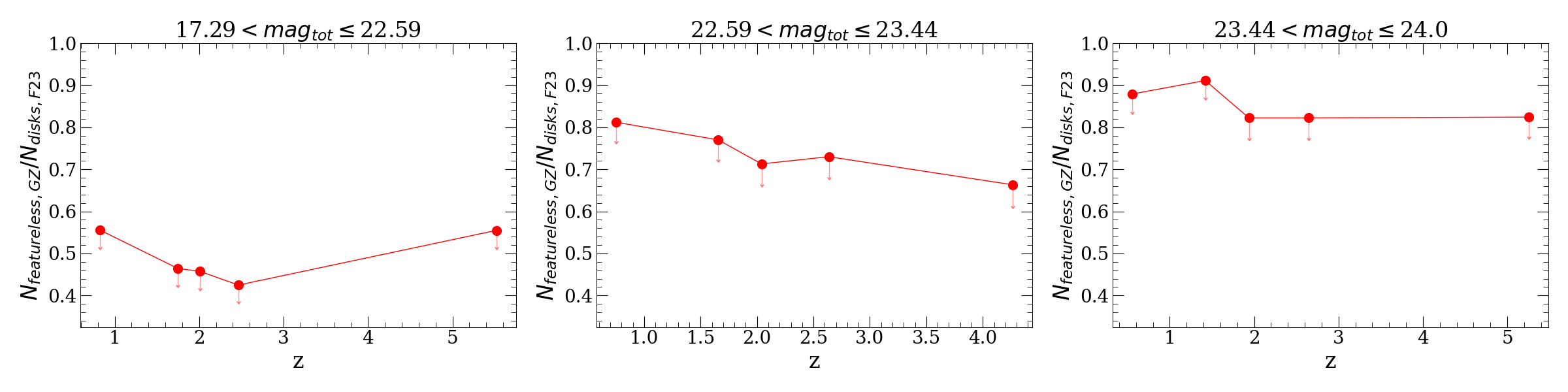}
\includegraphics[width=\textwidth]{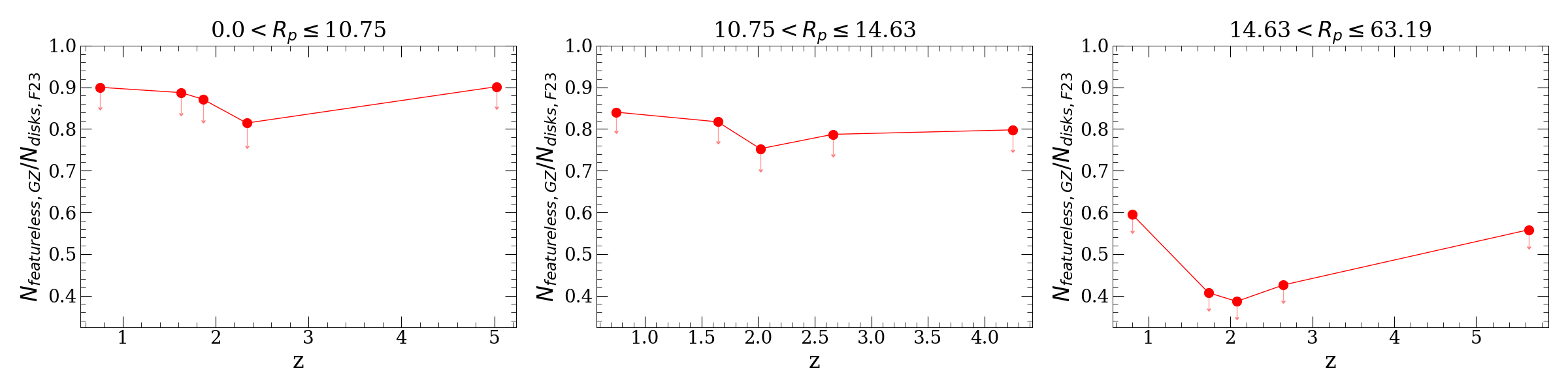}
\caption{The upper limits on the fraction of apparently featureless discs identified by GZ CEERS compared to F23 as a function of redshift, as in Figure~\ref{fig:featurelesswithz}, split into three even magnitude bins (top) and galaxy Petrosian radius bins (bottom). Upper limit arrows are scaled to show the uncertainties on each calculated fraction. Featureless discs are identified as those with GZ CEERS vote fractions, $p_{\rm{feat,GZ}}<0.3$. Whereas \protect\citetalias{ferreira23} visually identify discs as those galaxies with an expert disc vote fraction $p_{\rm{disc,F23}}>0.5$. The inverse of this measure is the featured disc vote fraction, $p_{\rm{feat,GZ}}\geq0.3$, shown by the secondary y-axis on the right. The fraction of featureless discs increases from $\sim40-60\%$ for the brighter, larger galaxies (right panels), to $80-90\%$ for the fainter, smaller galaxies (left panels) at all redshifts.}
    \label{fig:magandradiusbinsz}
\end{figure*}

\begin{figure}
\includegraphics[width=0.5\textwidth]{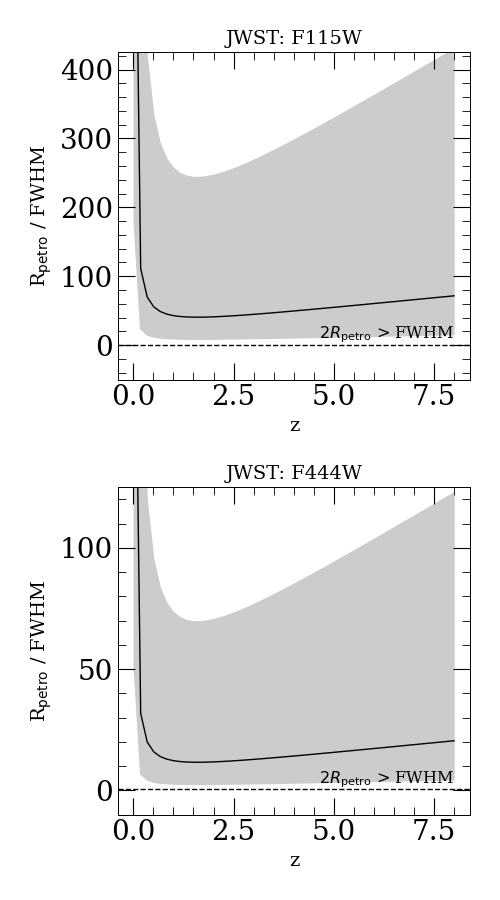}
    \caption{The Petrosian radii compared to the FWHM of a given JWST filter for an artificially redshifted population of disc galaxies with redshift. The Petrosian radii distribution is from the distribution of discs identified in GZ CEERS. The median of the population at each redshift is shown by the black solid line, and the minimum and maximum shown by the shaded region for both of the F115W (top) and F444W (bottom) filters. \rjs{The detection limit for each filter at $2R_{\rm{petro}}$ > FWHM is shown by the horizontal dashed line in each panel}.}
    \label{fig:detectionlimits}
\end{figure}

In Figure~\ref{fig:featurelesswithz}, we showed how $\sim58-76\%$ of discs in CEERS imaging were featureless using a combination of volunteer and expert classifications from GZ CEERS and \protect\citetalias{ferreira23} respectively. The first question we consider is whether these discs are truly featureless in morphology or whether the resolution limit of JWST results in an inability to classify any features. Although a marked improvement between JWST and HST was seen in Figure~\ref{fig:fractionwithz}, it is possible that features are still being missed in JWST imaging due to observational effects. For example, in Figure~\ref{fig:magandradiusbinsz}, we show how the featureless fraction with redshift changes when controlling for total galaxy magnitude, $\rm{mag}_{\rm{tot}}$ (across all CEERS bands used to create images shown to GZ volunteers), and Petrosian radius, $R_p$. The fraction of featureless discs increases from $\sim40-60\%$ for brighter, larger galaxies (right panels of Figure~\ref{fig:magandradiusbinsz}), to $80-90\%$ for fainter, smaller galaxies (left panels of Figure~\ref{fig:magandradiusbinsz}) at all redshifts. The featureless disc fraction in CEERS is therefore obviously affected by observational biases. 

In order to quantify whether this is first due to the resolution limit of JWST, we consider the typical size of morphological features compared to the FWHM of the shortest (F115W; $0.040\arcsec$) and longest (F444W; $0.145\arcsec$) waveband filters used to produce the false colour NIRCam imaging shown to volunteers in the GZ CEERS project. We sourced FWHM values of the empirical PSF of each filter, measured using observations of the standard star P330E from the Cycle 1 Absolute Flux Calibration program\footnote{\url{https://jwst-docs.stsci.edu/jwst-near-infrared-camera/NIRCam-performance/NIRCam-point-spread-functions\#NIRCamPointSpreadFunctions-PSFFWHM}}. 

We begin by considering the Petrosian radius of each galaxy (measured by \citetalias{ferreira23} during their parametric morphology fits), and artificially redshift the entire population to consider their apparent size in bins from $0<z<8$. This is shown in Figure~\ref{fig:detectionlimits} with the median of the population shown by the black solid line, and the minimum and maximum shown by the shaded region for both of the F115W and F444W filters. The detection limit at $2R_{\rm{petro}}$ > FWHM is shown by the dashed line. The effect of the turnover in the angular diameter distance at $z\sim1.5-2$ can be seen, which we also suggest may partially explain our results in Figures~\ref{fig:featurelesswithz} and~\ref{fig:fractionwithz}. We find that the entire population of disc sizes are visible above the detection limit in each of the JWST filters at all the artificial redshifts investigated, suggesting that we should indeed be able to detect all the discs at each redshift probed. However, this is the full size of the disc of the galaxy; features such as bars and spiral arms will be smaller than this. It is only when we investigate features which are $<0.1R_{\rm{petro}}$ that a small fraction of the population ($\sim1\%$) drops below the FWHM resolution limits of the F444W filter. Given that the typical sizes of bulges, bars, and spiral arms in galaxies are a significant fraction of the radius of the galaxy \citep[at least at lower redshifts, e.g.][showed how the ratio between bar radius and galaxy radius stayed constant at $\sim0.5$ at $z<0.84$]{Kim2021}, the resolution of JWST is therefore not a major limiting factor for detecting features. This is in agreement with the findings of \citet{Liang2024} who found that bar detections remain effective at $\sim100\%$ as long as the intrinsic bar size remains at $>~2~\times~$FWHM.

However, our investigations on disc and feature size above does not take into account the surface brightness of these features. \citet{Liang2024} did a thorough investigation of a set of simulated CEERS images (using the method outlined in \citealt{Yu2023}), taking into account various observational effects including spectral change, cosmological surface brightness dimming, luminosity evolution, physical disc size evolution, the shrinking in angular size due to distance, the decrease in resolution, and increase in signal-to-noise ratio with redshift. They find that as the redshift increases to $z=3$, the effectiveness of detecting bars gradually decreases. It declines from $100\%$ of bars successfully identified at $z\sim0$, to approximately $55\%$ of bars successfully identified at $z\sim3$, suggesting that up to half of all bars are missing at higher redshift due to the impact of observational effects listed above. A similar behaviour has been found in simulated images for HST and JWST for spiral arms \citep{Bentabol2022, Kuhn2024}. 

We can therefore expect at least half of the featureless discs that we have identified do indeed have features which have been missed due to observational effects such as surface brightness dimming. Conversely, this suggests that at most, half of our sample truly are featureless discs. However, this result is challenged by the findings of \citet{Vega-Ferrero2024} who find that half the galaxies in CEERS classified as discs (either using machine learning or expert visual classifications) populate a similar region of their machine learning algorithm's representation space as simulated galaxies from TNG50 \citep{Nelson2019} which have low stellar specific angular momentum and non-oblate structure. 

The upper limit on the apparently featureless fraction seen in Figure~\ref{fig:featurelesswithz} of $\sim58-76\%$ is clearly an overestimate. Given the findings of \citet{Yu2023, Liang2024, Kuhn2024, Vega-Ferrero2024} and the results discussed above in Figures~\ref{fig:detectionlimits} \& \ref{fig:magandradiusbinsz}, we revise our upper limits to suggest that \emph{at most} half of our sample are truly featureless discs. \rjs{We therefore halve the upper limit fractions found in Figure~\ref{fig:featurelesswithz} ($\sim58-76\%$) to estimate that the featureless disc fraction in JWST CEERS imaging is more likely to have an upper limit in the range of $\sim29-38\%$.}

\subsection{Are these truly dynamically cold discs?}\label{subsec:colddiscs}

Without the features that identify a galaxy as rotationally supported (e.g. bars, spiral arms), identifying an apparently featureless galaxy as a truly dynamically cold system with imaging data alone is not trivial. However, the light profile of a galaxy is thought to trace whether a galaxy is rotational or dispersion dominated. An exponential light profile (with Sérsic index, $n\approx1$) is consistent with a rotationally supported disc \citep{Sersic1968, Trujillo2001, Elmegreen2013, Vika2015}. \citet{Elmegreen2013} showed how galaxies with exponential light profiles with no rotation are unstable; therefore, if a galaxy has a Sérsic index, $n\approx1$, then we can assume it is a dynamically cold system. Note that the relevance of the Sérsic fitting function out to such high redshifts has yet to be established, however \citet{Yu2023} showed with simulated CEERS images, that the average difference between the measured Sérsic index, $n$, and the true Sérsic index, $n_{True}$ is small, at \protect{$n_{True}-n\sim-0.03$}; suggesting that Sérsic fitting can be successfully applied without significant bias for galaxies observed in CEERS. We can therefore use the light profile fits from \citetalias{ferreira23}, specifically the two-dimensional Sérsic indices, to probe whether the featureless discs identified here are truly discs. 

\begin{figure}
\includegraphics[width=0.5\textwidth]{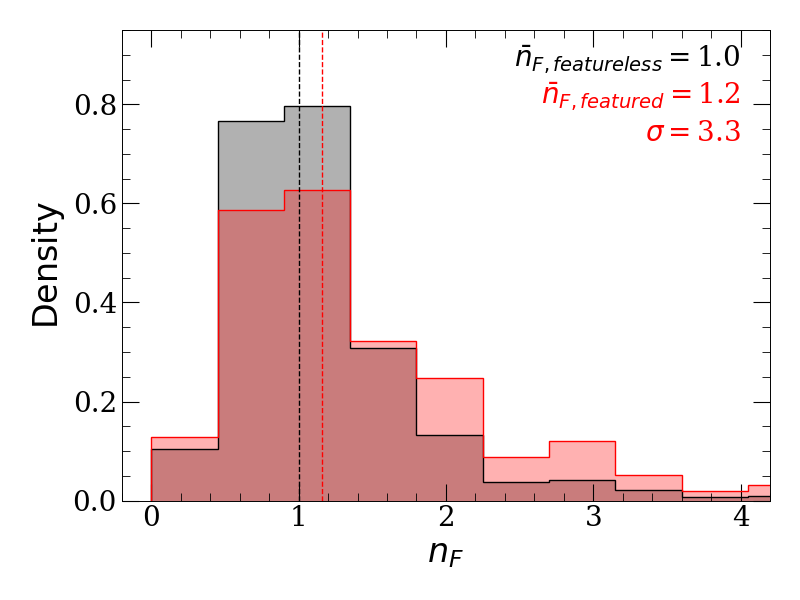}
    \caption{The distribution of the Sérsic index, $n_{F}$, from \protect\cite{ferreira23} for the featureless (black histogram) and featured (red histogram) disc galaxy samples. The median Sérsic fit, $\bar{n}_{F}$, for each sample is shown in the top right corner, along with the significance, $\sigma$, of an Anderson Darling test as to whether the two distributions are statistically distinguishable.}
    \label{fig:sersic}
\end{figure}

We show the distribution of Sérsic index, $n_{F}$, for the apparently featureless (black histogram) and featured (red histogram) disc samples in Figure~\ref{fig:sersic}. The median Sérsic index for the featureless disc sample is $\bar{n}_{F}=1.0$ (with $80\%$ of the sample with $n_{F}<1.5$), and for the featured sample is $\bar{n}_{F}=1.2$ (with $64\%$ of the sample with $n_{F}<1.5$). The two distributions are statistically significantly different from each other, with an Anderson-Darling test \citep{AndersonDarling1954} resulting in $\sigma=3.3$; note that Sérsic fits are typically offset by the presence of features. Despite this difference, the two samples both have average light profiles consistent with an exponential profile (as shown in Figure~\ref{fig:sersic}), i.e. rotationally supported, suggesting the apparently featureless systems identified here out to $z=7.35$ are indeed dynamically cold. Future studies using follow-up spectroscopic data of the CEERS footprint with NIRSpec would be able to determine the velocity dispersion and kinematics (along with obtaining spectroscopic redshifts) of the apparently featureless disc systems identified here and confirm if they are truly rotationally supported.

\subsection{Why do these discs lack features?}\label{subsec:lackfeatures}
For these galaxies to be truly featureless (lacking spiral arms, bars, bulges, clumps), there must be no density waves, disc instabilities, tidal interactions, or other processes which induce perturbations in the stellar disc (note that \citealt{Elmegreen2013} show how exponential discs develop from stellar scattering alone, and no other perturbations are required). This could be the case if they are incredibly gas rich; mergers of galaxies with high gas fractions are seen in simulations to reform a disc post-merger \citep{SparreSpringel2017,Bois2011} and gas rich systems are more resistant to buckling and instabilities \citep{Lokas2020}. Merger rates are known to increase with redshift, at least out to $z\sim6$ in HST CANDELS  \citep[e.g.][]{Duncan2019}, out to $z\sim8$ followed by a flat evolution in JWST Cycle-1 fields \citep{Duan2024}, and in simulations \citep{Rodriguez-Gomez2015, Snyder2019}. Investigations on merger rates with JWST at higher redshift are still forthcoming. Similarly, \cite{Walter2020} showed how the ratio of molecular gas to stellar mass density increases with redshift out to $z\sim4$ in deep volumetric surveys. In addition, \cite{Wang2022} use ALMA data to show how the mean molecular gas fraction increases out to $z\sim4$ for all stellar masses. While we are wary of extrapolating these trends beyond $z\sim4$, more recent JWST studies of individual systems in conjunction with ALMA have supported the assumption that gas fraction increases with redshift. For example, \cite{Heintz2023} find a molecular gas fraction of $f_{\rm{gas}} = M_{\rm{gas}}/(M_{\rm{gas}} + M_{*})>90\%$ in a galaxy at $z=8.496$.  

However, \cite{Fujii2018, JBH2023, JBH2024} show in their high resolution hydrodynamical simulations how gas fraction is secondary to the baryonic (i.e. stellar plus gas) vs. dark matter fraction within the disc, \protect{$f_{\rm{disc}}~=~\left(\frac{V_{\rm{c,disc}}(R_s)}{V_{\rm{c,tot}}(R_s)}\right)^2$}, where $R_s=2.2 R_{\rm{disc}}$, with bar length, mass, and formation time all scaling inversely with $f_{\rm{disc}}$. They find that at low $f_{\rm{disc}}\lesssim0.3$, a bar does not form in a stellar disc; this remains true even at high gas fractions. For a baryon dominant disc, i.e. $f_{\rm{disc}}>0.5$, they find that at all gas fractions an intermittent bar forms within $\sim500$~Myr. For context, at $z=0$, baryonic matter dominates over dark matter within the solar radius of the Milky Way, with $f_{\rm{disc}}\sim0.5-0.65$ \citep{JBHG2016}. Out to higher redshifts, \cite{Genzel2020, Price2021} have shown that baryons also dominate over dark matter within discs between $z\sim1-2.5$. Therefore, the apparently featureless disc galaxies found in this study in CEERS imaging might be galaxies which are still dominated by dark matter within the disc, with as yet insufficient time to build up their baryonic mass fraction.

Given this prevalence of increased merger rates, gas fractions, and baryonic disc fractions at increasing redshift, it follows that discs could be forming and re-forming at all epochs leading to featureless discs. \citet{Simmons2014} also highlight that such an environment with dynamically warm discs, would lead to the formation of short-lived bar structures at high redshift. We would therefore expect a low \citep[][estimate a bar fraction of $\sim10\%$]{Simmons2014}, yet stable, bar fraction in the GZ CEERS sample in a similar redshift range, which we investigate in a forthcoming paper \citep{geron24prep}.

In order to confirm whether CEERS galaxies are indeed gas-rich or baryon dominated, more follow-up will be required combining the power of JWST and ALMA across a morphologically diverse sample of galaxies, including a large number of apparently featureless discs. 

\section{Conclusions}\label{sec:conclusions}

JWST has allowed the identification of morphological features at higher redshifts than ever before, such as spiral arms \citep[out to $z\sim4$, e.g.][]{Wu2023, Kuhn2024}, bars \citep[out to $z\sim4$, e.g.][]{Guo2023, Constantin2023, LeConte2024} and clumps \citep[out to $z\sim8$, e.g][]{Kalita2024,Tohill2024}. However, such studies which identify solely large-scale features in an attempt to characterise the epoch of the emergence of the Hubble Sequence do not account for the possibility of a featureless disc, and so will underestimate the redshift at which discs emerge. Thankfully, the Galaxy Zoo collaboration has completed a pilot study using public JWST CEERS \citep{Finkelstein2023} NIRCam imaging of $\sim7000$ galaxies \citep{masters24prep}, which specifically asks users if a galaxy is featured or otherwise. The GZ CEERS classifications in combination with expert visual or parametric morphological classifications, are therefore incredibly useful for identifying a sample of this forgotten population of featureless disc galaxies at high redshift.

In this study we combined GZ CEERS vote fractions with Sérsic index estimates and expert visual morphologies from \citet{ferreira23} to identify a sample of apparently featureless discs (i.e. a mix of truly featureless discs, and discs with features too faint to be visible at these redshifts). From visual inspection we show the highest redshift apparently featureless disc we identified is CEERS-21290 at $z_{\rm{phot}}=7.4$ (with Sérsic index, $n=1.5$; see Figure~\ref{fig:highestz}), and similarly, the highest redshift featured disc we identified is CEERS-11131 with both disc and bulge feature at $z_{\rm{phot}}=5.5$ (with Sérsic index, $n=1.1$; see Figure~\ref{fig:highestz}). Spectral follow-up will be required to confirm the redshifts of these systems, however, it is clear that any study of the epoch of the emergence of discs in the Universe's history must therefore include featureless discs, which can be identified out to much higher redshift than featured discs. Our apparently featureless discs sample therefore allows us to put limits on the redshift epoch at which discs, and the features which define the Hubble Sequence, emerged, which will help inform the next generation of cosmological and hydrodynamical simulations. Our conclusions are as follows:
\begin{enumerate}
\item{We investigate the upper limit on the apparently featureless disc fraction as a function of redshift, finding a range between $58-76\%$ between $0<z_{\rm{phot}}<7.4$ (see Figure~\ref{fig:featurelesswithz}). However, we know this will be an overestimate due to observational biases; the visibility of features will naturally decline with redshift. Therefore, we first discussed in Section~\ref{subsec:zlimit} whether these galaxies are truly featureless. Using simulated CEERS imaging from previous studies \citep{Yu2023, Liang2024, Kuhn2024} we discuss how this is likely an overestimate given observational effects, and estimate that the truly featureless disc fraction in JWST CEERS imaging is at most half of the fraction reported in Figure~\ref{fig:featurelesswithz}, with an upper limit in the range $29-38\%$.}

\item{Regardless of whether they are featureless or not, we then discussed in Section~\ref{subsec:colddiscs} whether these galaxies are truly dynamically cold discs. The distribution of Sérsic indices (shown in Figure~\ref{fig:sersic}) for both the featureless and featured disc galaxy samples peak at $n_F\sim1$, suggesting that these truly are dynamically cold systems. This suggests that the Hubble Sequence was therefore in place from at least $z_{\rm{phot}}\sim7.4$.} 

\item{Assuming therefore that our sample indeed contains truly featureless discs, in Section~\ref{subsec:lackfeatures} we discuss why these systems lack features. Our hypothesis is two fold. Firstly, that the lack of features in these disc systems is due to a higher gas fraction which allows the discs to reform. Future observations with ALMA and JWST NIRSpec should be able to reveal their gas content and determine the true kinematic nature of these featureless systems. Second, that the baryonic vs. dark matter fraction, $f_{\rm{disc}}$, in the high redshift featureless systems is too high for features to have yet developed, i.e. dark matter still dominates over baryonic matter in these systems. Hydrodynamical zoom-in simulations with sufficient particle resolution predict that structure formation from instabilities, such as bars, will not occur until $f_{\rm{disc}}\gtrsim0.5$ \citep{Fujii2018, JBH2023, JBH2024}.}
\end{enumerate}

The current generation of cosmological volume simulations employing $\Lambda$CDM predict disc formation at redshifts $z\sim1-4$ \citep{Parry2009, Dubois2016, Pillepich2019, Semenov2024b}. This is in direct conflict with the observational results presented here showing that while featured discs are identifiable out to $z_{\rm{phot}}=5.5$ ($t_{\rm{age}}\sim1~\rm{Gyr}$), apparently featureless discs are identifiable out to $z_{\rm{phot}}=7.4$ ($t_{\rm{age}}\sim0.7~\rm{Gyr}$) in JWST CEERS imaging. 

\section*{Acknowledgements}
The authors would like to thank the referee for a thoughtful report on our paper, along with Joss Bland-Hawthorn for an excellent discussion on the implications of this work.  RJS would also like to highlight the contributions of her cat Pip to this work, who took to the keyboard a number of times during preparation of this manuscript; sadly, most of Pip's contributions had to be removed. 

This research was supported by the International Space Science Institute (ISSI) in Bern, through ISSI International Team project \#23-584 (Development of Galaxy Zoo: JWST). RJS gratefully acknowledges support from the Royal Astronomical Society Research Fellowship. BDS acknowledges support through a UK Research and Innovation Future Leaders Fellowship [grant number MR/T044136/1]. MW is a Dunlap Fellow and TG is a Rubin Fellow at the Dunlap Institute; the Dunlap Institute is funded through an endowment established by the David Dunlap family and the University of Toronto.  LFF, HR and KBM acknowledge partial support from the US National Science Foundation under grants IIS 2006894, IIS 2334033 and NASA award \#80NSSC20M0057. ILG has received the support from the Czech Science Foundation Junior Star grant no. GM24-10599M. 

This work is based on observations made with the NASA/ESA/CSA James Webb Space Telescope. The data were obtained from the Mikulski Archive for Space Telescopes at the Space Telescope Science Institute, which is operated by the Association of Universities for Research in Astronomy, Inc., under NASA contract NAS 5-03127 for JWST. These observations are associated with ERS program \#1345. This work is in part based on observations taken by the CANDELS Multi-Cycle Treasury Program with the NASA/ESA HST, which is operated by the Association of Universities for Research in Astronomy, Inc., under NASA contract NAS5-26555.

\section*{Data Availability}

 All data used in this paper is available from the Mikulski Archive for Space Telescopes at the Space Telescope Science Institute, or upon request to the first authors.



\bibliographystyle{mnras}
\bibliography{references} 








\bsp	
\label{lastpage}
\end{document}